\documentstyle[twoside,fleqn,espcrc2]{article}
\newcommand{\be}{\begin{equation}}
\newcommand{\ee}{\end{equation}}
\newcommand{\bea}{\begin{eqnarray}}
\newcommand{\eea}{\end{eqnarray}}

\newcommand{\half}{{\scriptstyle{{1\over 2}}}}

\newcommand{\quart}{{\scriptstyle{{1\over 4}}}}
\newcommand{\real}{\relax{\rm I\kern-.18em R}}
\newcommand{\zahlen}{{\rm Z \!\! Z}}

\newcommand{\Tr}{\mbox{\,Tr\,}}

\newcommand{\Pexp}{{\rm Pexp}}

\newcommand{\flt}{{\rm flat}}
\newcommand{\mod}{{\,\rm mod\,}}
\newcommand{\ctg}{{\rm ctg}}
\newcommand{\coker}{{\rm coker\,}}

\newcommand{\cP}{{\cal P}}
\newcommand{\cO}{{\cal O}}
\newcommand{\cd}{{\hat d}}
\newcommand{\cE}{{\cal E}}
\newcommand{\ch}{{\rm ch}}

\newcommand{\tr}{\mbox{\,tr\,}}

\newcommand{\veps}{\varepsilon}
\newcommand{\fie}{\varphi}

\newcommand{\integer}{\relax{\rm I\kern-.18em N}}

\newcommand{\cN}{{\cal N}}
\newcommand{\cM}{{\cal M}}

\newcommand{\phd}{{\vphantom{\dagger}}}

\newcommand{\AmS}{{\protect\the\textfont2
  A\kern-.1667em\lower.5ex\hbox{M}\kern-.125emS}}
\title{
\vskip-1cm\hfill{\tenrm INLO-PUB-24/95}\vskip5mm
Instanton Moduli for $T^3\times\real$}
\author{Pierre van Baal
\address{Instituut-Lorentz for Theoretical Physics,\\
University of Leiden, PO Box 9506,\\
NL-2300 RA Leiden, The Netherlands.}
\thanks{Based on a talk presented at the 29th International Symposium
on the Theory of Elementary Particles (Buckow, 29 August - 2 September, 1995)}}
\begin{document}

\begin{abstract}
We review the specific problems that arise when studying instantons on a torus.
We discuss how the Nahm transformation shows that no exact charge one instanton
on $T^4$ can exist. However, taking one of the directions (the time) to
infinity, it can be shown that vacuum to vacuum tunnelling solutions exist.
A precise description of the moduli space for $T^3\times\real$, studied
numerically using lattice techniques, remains an interesting open problem.
New is an explicit application of the Nahm transformation to (anti-)selfdual
constant curvature solutions on $T^4$ and a discussion of its properties
relevant to instantons on $T^3\times\real$.
\end{abstract}
\maketitle

\section{Introduction}

Instantons and monopoles have become important mathematical tools to study
invariants of four dimensional manifolds~\cite{don,wit}. This is due to the
fact that their moduli space (the parameter or solution space) depends on
the space-time on which the self-duality equations are studied. To some
extent we can compare~\cite{top} the situation with the study of harmonic
forms on a manifold and its relation to the DeRham cohomologies. Like in that
case, also here there are certain spaces for which there are no solutions to
the equations studied.

The torus $T^4$ turns out to be an example where the moduli space for
instantons of unit charge is empty. For higher charges existence was proved
by Taubes~\cite{tau} quite some time ago. The general proof for existence
takes a small localized instanton from $\real^4$ that is matched smoothly
to the flat connection of the trivial bundle. In particular if the moduli
space of flat connections has non-zero dimension, there can be obstructions
against this procedure. This generally occurs when $H_2(M,\zahlen)$ is
non-trivial, as the trace of the Wilson loop for a flat connection
can be non-trivial when the loop cannot be contracted to a point.

Using the Nahm transformation~\cite{nam}, it can be proven that indeed charge
one instantons do not exist for $T^4$. The Nahm transformation will be
reviewed in the next section (see also refs.~\cite{cor}). It maps
self-dual solutions on $T^4$ to self-dual solutions on
the dual
torus~\cite{lec,sch}. This map is an involution, i.e. its square is the
identity, and it preserves metric and hyperK\"ahler structures of the moduli
spaces~\cite{bra}.

{}From the physical point of view it is somewhat disturbing, in particular
when scaling up the volume to sizes much bigger than $1/\Lambda_{\rm QCD}$,
that the existence would depend on the geometry of space-time. The point is
of course that one can get arbitrarily close to a solution, for sufficiently
large volumes, but an exact solution is only achieved in the singular limit.
Nevertheless, the obstruction against the existence of a solution is
relatively mild. For twisted boundary conditions~\cite{tho}, the existence of
charge one instantons can be understood from the fact that any twist removes
the continuous degeneracy in the moduli space of (four dimensional) flat
connections, removing the obstruction to glueing in a localized
instanton~\cite{mac}. Twisting the boundary conditions in the time
direction provides the proper framework for understanding the existence of
charge one instanton solutions on $T^3\times\real$, as was confirmed
by numerical lattice studies~\cite{oim}.

The physical picture is most simply explained at infinite time. To keep the
action of the imaginary time solution finite, the magnetic (and electric)
energy should vanish at either end. This means that the connections at
these ends are flat connections on $T^3$, whose moduli space is an
orbifold (for SU(2) it is $T^3/Z_2$), most conveniently parametrized by
the trace of the Wilson loops along the three generators of the three-torus.
When twisting in the time direction, at least one of these Wilson loops is
required to have opposite signs at the two ends, somewhat misleadingly this
can be compared to anti-periodic boundary conditions. Apparently, instanton
solutions are not compatible with periodic boundary conditions. From the
Hamiltonian point of view, in the $A_0=0$ gauge, there is no difference
between the twisted and periodic case, except that in the periodic case the
trace of the Polyakov loops at $t=\pm\infty$ are identical and for the twisted
case opposite in sign. In particular the sphaleron giving the saddle point,
or minimal barrier that separates two vacua, is obtained by solving the static
Yang-Mills equations on the (untwisted) three-torus~\cite{mar}. This sphaleron
would not exist in an infinite volume, but the finite volume breaks the
classical scale invariance, which also implies that the scale parameter
of the instanton is no longer associated to a symmetry, but it is still
a non-trivial moduli parameter.

An interesting open problem remains if there are solutions on $T^3\times\real$
that can not be compactified to $T^4$, with twist in the time.

\section{The Nahm transformation}

It is convenient to view $T^4$ as $\real^4/\Lambda$, where
$\Lambda=\oplus_{\mu=1}^{4}\zahlen e_\mu$ is a four dimensional lattice.
The connection one-forms $\omega(x)=A_\mu(x)dx_\mu$ are invariant up to
a gauge transformation under translation over a lattice vector. These
gauge transformations, also called cocycles, satisfy cocycle conditions
so as to assure one has an appropriate principal fiber bundle over the
torus:
\bea
\omega(x+\lambda)&=&g_\lambda(x)(\omega(x)+d)g^{-1}_\lambda(x),\nonumber\\
g_{\lambda+\mu}(x)&=&g_\lambda(x+\mu)g_\mu(x),\quad\lambda,\mu\in\Lambda.
\eea
We will consider anti-selfdual connections with gauge group SU($N$) whose
curvature satisfies
\be
\Omega=d\omega+\omega\wedge\omega=-{}^*\Omega.
\ee
In local coordinates the curvature is given in terms of the field strength
by $\Omega\!=\!\half F^{\mu\nu}dx_\mu\wedge dx_\nu$, whereas ${}^*\Omega$ is
defined similarly in terms of the dual of the field strength,
$\tilde F^{\mu\nu}=\half\veps^{\mu\nu\lambda\sigma}F_{\lambda\sigma}$.
The topological charge $k$ is given by the integral over the second Chern
class,
$k=\int\Tr\left(\frac{\Omega}{2\pi i}\wedge\frac{\Omega}{2\pi i}\right)$.

The Nahm transformation, $\hat\omega\equiv\cN\omega$, will define a connection
on the dual torus $\hat T^4=\real^4/\hat\Lambda$, where
\be
\hat\Lambda=\{\mu\in\real^4|<\mu,\lambda>\in\zahlen,\ \forall\lambda\in
\Lambda\}\ee
is the lattice dual to $\Lambda$. We will show that $\hat\omega$ is
a SU($k$) anti-selfdual connection with topological charge $N$. The rank
of the gauge group and the topological charge are therefore interchanged under
the Nahm transformation. If we denote by $\cM_{N,k}$ the moduli space
of SU($N$) charge $k$ instantons, the Nahm transformation induces
a map between moduli spaces, $\cN:\cM_{N,k}\!\rightarrow\!\cM_{k,N}$,
which is an involution that preserves the natural metric and hyperK\"ahler
structure of the moduli space~\cite{bra}. The dimension of the moduli
space, $4Nk$, is indeed symmetric under interchanging $k$ and $N$.

We will now explain the essential ingredients of the Nahm transformation.
The starting point is that a charge $k$ (anti-)instanton has $k$ (negative)
positive chirality zero-modes for the massless Dirac equation, which is most
conveniently written in the Weyl format. We introduce the four unit
quaternions $\sigma_\mu$, with $\sigma_0=1_2$ and $\sigma_i=-i\tau_i$, where
$\tau_i$ are the usual Pauli-matrices. The Weyl operators are
given by $D^-\equiv D$ and $D^+\equiv-D^\dagger$, with
\be
D\equiv\sigma_\mu D_\mu(A)=\sigma_\mu(\partial_\mu+A_\mu).
\ee
Hence, in the background of a charge $k$ instanton there are $k$
independent solutions to $D\Psi=0$. For $\Psi(x)$ to be defined as a two-spinor
on the torus one requires $\Psi(x+\lambda)=g_\lambda(x)\Psi(x)$. One now
adds a spectral parameter $z_\mu\in\real^4$ in the form of a flat abelian
connection
\be
\omega_z=\omega+2\pi iz_\mu dx^\mu,
\ee
which leaves the curvature unchanged, $\Omega_z=\Omega$, and in particular
anti-selfdual. Hence there is a smooth family of $k$ fermionic zero-modes
\be
D_z\Psi_z^{(i)}(x)\!=\!\sigma_\mu(\partial_\mu\!+\!A_\mu\!+\!2\pi i z_\mu)
\Psi_z^{(i)}(x)\!=\!0.
\ee
{}From this family one construct a connection $\hat\omega$ by
\be
\hat A_\mu^{ij}(z)=\int_{T^4}\!\!d_4x~\Psi_z^{i}(x)^\dagger\frac{\partial}{
\partial z_\mu}\Psi_z^{(j)}(x).
\ee
This can be seen to form a connection on the dual torus using the fact
that the $\Psi_z^{(i)}(x)$ form a complete orthogonal set of solutions
of the Weyl equation and the observation that
$e^{-2\pi ix\cdot\hat\lambda}\Psi_z^{(i)}(x)\in\ker\,
D_{z\!+\!\hat\lambda}$,
such that
\be
\Psi_{z\!+\!\hat\lambda}^{(i)}(x)=e^{-2\pi ix\cdot\hat\lambda}\Psi_z^{(j)}(x)
S^{ji}_{\hat\lambda}(z),
\ee
with $S_{\hat\lambda}(z)$ a unitary $k\times k$ matrix, which defines the
(inverse) of the cocycle for $\hat\omega\equiv\hat A_\mu(z)dz_\mu$
as a connection on $\hat T^4$
\be
\hat A_\mu(z+\hat\lambda)=S^{-1}_{\hat\lambda}(z)(\hat A_\mu(z)+\frac{
\partial}{\partial z_\mu})S_{\hat\lambda}(z).
\ee
We can use the Atiyah-Singer family index theorem~\cite{ati} to relate
the Chern character of the bundle $E$, associated with the connection $\omega$
to the Chern character of the bundle $\hat E$, associated with the connection
$\hat\omega$. For this it is necessary to assume that $\omega$ is
1-irreducible or WFF (without flat factors~\cite{don}), which is equivalent
to stating that $\coker D_z\!=\!\ker\, D^+_z\!=\!0$, implying that
$\hat E_z\!\equiv\!\ker\, D_z$ is smooth (remember that the index theorem
states that $k=\dim\,\ker\, D^--\dim\,\ker\, D^+$). One can view $\omega_z$
as a connection over $T^4\times\hat T^4$, where the abelian part
has a curvature $2\pi idz_\mu\wedge dx_\mu$ (forming the so-called Poincar\'e
bundle $\cP$~\cite{bra}). It now follows that
\be
\ch(\hat E)=\int_{T^4}\ch(E)\wedge\ch(\cP).
\ee
This is easily seen to interchange the rank and topological charge between
the original and Nahm bundles. Also the first Chern classes will be related,
but we will assume that $E$ is a SU($N$) bundle, for which the
first Chern class vanishes. The family index theorem guarantees that the
Nahm bundle $\hat E$ also has vanishing first Chern class, from which
it follows that $\hat E$ is a SU($k$) bundle with topological charge $N$.
If, however, the first Chern class does not vanish (in which case the
topological charge should be determined from the first Pontryagin class),
one has~\cite{bra}
\be
c_1(\hat E)=-\int_{T^4}(dz_\mu\wedge dx_\mu)^2\wedge c_1(E).
\ee

A direct corollary is now that charge 1 instantons cannot exist
on $T^4$. Suppose they would exist. The Nahm bundle would give rise to a
U(1) bundle of charge $N$, which is impossible as the first Chern
class vanishes and U(1) bundles have always vanishing second Chern class.

The family index theorem only provides topological information on the
Nahm bundle obtained from an anti-selfdual connection $\omega$. To
demonstrate that $\hat\omega$ is also an anti-selfdual connection, a little
more work is needed. As some of the ingredients will be important for
the formulation of the Nahm transformation on $T^3\times\real$ we provide
a few details necessary to understand this beautiful result due to Nahm,
who originally introduced his transformation for the study of monopoles,
as a generalization~\cite{nam,cor} of the ADHM construction~\cite{adhm}.
A crucial ingredient is formed by the Weitzenb\"ock formula
\be
D^-_zD^+_z=D^2_\mu(A_z)+\sigma^\phd_{[\mu}\sigma^\dagger_{\nu]}
F_{\mu\nu},
\ee
using $\sigma_\mu\sigma_\nu^\dagger=\delta_{\mu\nu}+\sigma^\phd_{[\mu}
\sigma^\dagger_{\nu]}$. Since $\sigma^\phd_{[\mu}\sigma^\dagger_{\nu]}
\equiv\sigma_i\eta^i_{\mu\nu}$ is a selfdual tensor,
we see that $D^-_zD^+_z=D^2_\mu(A_z)$. Its kernel is trivial
for $\omega$ WFF and the Greens function $G_z=(D_z^-D_z^+)^{-1}$
commutes with the quaternions $\sigma_\mu$.

It is now remarkably simple to show that $\hat\omega$ is also anti-selfdual
\be
\hat\Omega^{ij}(z)=\cd\hat\omega+\hat\omega\wedge\hat\omega
=<\cd\Psi_z^{(i)}|1-P|\cd\Psi^{(j)}_z>,
\ee
where
\be
\cd\equiv dz_\mu\frac{\partial}{\partial z_\mu},\quad
P\equiv |\Psi_z^{(k)}><\Psi_z^{(k)}|.
\ee
One easily shows that
\bea
P&=&1-D_z^+G_zD_z,\\ D_z\cd\Psi_z^{(i)}&=&[D_z,\cd]\Psi_z^{(i)}=
-2\pi i\sigma_\mu\Psi_z^{(i)}dz_\mu,\nonumber
\eea
such that
\be
\hat F_{\mu\nu}^{ij}(z)=8\pi^2<\Psi^{(i)}_z|\sigma^\dagger_{[\mu}
\sigma^\phd_{\nu]}G_z|\Psi_z^{(j)}>.
\ee
There are two essential ingredients that enter these manipulations. First,
on $T^4$ we can perform partial integrations without picking up
boundary terms. Second, $G_z=(D^-_zD_z^+)^{-1}$ commutes with the
quaternions. Anti-selfduality immediately follows from the fact
that $\sigma^\dagger_{[\mu}\sigma^\phd_{\nu]}
\equiv\sigma_i\bar\eta^i_{\mu\nu}$
is an anti-selfdual tensor. When one or more space-time directions
become non-compact, boundary terms complicate the construction~\cite{nam,cor}.
For $T^4$, applying the Nahm transformation the second time (in case of
non-compact directions this requires modification), it can be shown that
$\cN^2\omega=\omega$, and the explicit form of $\hat\Psi_x(z)$ in terms of
$\Psi_z(x)$ allows one to show that metric and hyperK\"ahler structures
of the moduli spaces are preserved under $\cN$~\cite{bra}.

We note that in the case of twisted boundary conditions~\cite{tho}, it is not
possible to construct the Nahm transformation, as the fields need to be
invariant under the center of the gauge group, which is not the case for
the fermionic zero-modes required for the construction of $\hat\omega$.
Nevertheless, for gauge theories on $T^3\times\real$, where we have periodic
boundary conditions in the space directions and non in the time directions,
we can attempt to construct a variant of the Nahm transformation, as will
be discussed in the last section.

\section{Explicit example}

In general no explicit (anti-)selfdual connections on $T^4$ are known.
However, for some choices of the periods $\Lambda$ (or equivalently for
some choices of metrics), constant curvature solutions exist~\cite{tho2}.
A complete classification for SU(2) was given in ref.~\cite{con}. Under
deformations of $\Lambda$ these connections remain solutions, but are
no longer (anti-)selfdual. In absence of twist, that is as proper SU(2)
(rather than SU(2)/$Z_2$=SO(3)) bundles, their topological charge is always
even. They have U(1) holonomy, as there exists a gauge in which they are
abelian. Consequently they are examples of so-called reducible connections,
giving rise to singular points in the moduli space~\cite{don}. The term
reducible derives from the fact that these bundles decompose in the
direct sum of U(1) line bundles $E=L\oplus L^{-1}$, see ref.~\cite{don}.
As the Nahm transformation preserves the metric structure of the moduli
spaces, reducible connections should be mapped to reducible connections.
This implies that the Nahm transformation of a (anti-)selfdual constant
curvature connection is expected to be also a (anti-)selfdual constant
curvature connection, to be illustrated in this section with the help a
simple explicit example for SU(2) and topological charge 2.

We consider on $T^4=\real^4/\zahlen^4$ the connection
\be
A_\mu(x)=-\half\pi i n_{\mu\nu}x_\nu\tau_3,\quad n_{03}=n_{21}=2,
\ee
with the cocycles given by
\be
g_\lambda(x)=\alpha(\lambda)\exp(\lambda_\mu A_\mu(x)).
\ee
The so-called bi-characters $\alpha$ are given by
\be
\alpha(\lambda)\equiv\exp(-\half\pi i\sum_{\mu<\nu}\lambda_\mu n_{\mu\nu}
\lambda_\nu).
\ee
As in the study of the fluctuations around the constant curvature
connections~\cite{con}, the zero-modes (and for that matter the
whole spectrum) of the Weyl operators can be expressed in terms of
theta functions. This is most simply seen by introducing complex
coordinates
\bea
y_1=\frac{x_3+ix_0}{\sqrt{2}}&,&y_2=\frac{x_1+ix_2}{\sqrt{2}}\nonumber\\
u_1=\frac{z_3+iz_0}{\sqrt{2}}&,&u_2=\frac{z_1+iz_2}{\sqrt{2}}
\eea
and suitable creation and annihilation operators
\bea
a_i&=&-i\sqrt{2}\left(\frac{\partial}{\partial\bar y_i}+\pi(y_i+2iu_i)\right),
\nonumber\\
b_i&=&-i\sqrt{2}\left(\frac{\partial}{\partial y_i}+\pi(\bar y_i+2i\bar u_i)
\right),
\eea
which satisfy the commutation relations
\be
[a_i,a^\dagger_j]=4\pi\delta_{ij},\quad[b_i,b^\dagger_j]=4\pi\delta_{ij},
\ee
with all other commutators trivial. Hence all eigenfunctions can be
constructed, as for a harmonic oscillator, in terms of the functions
$\chi_z(x)$ and $\chi^*_{-z}(x)$, annihilated by respectively $a_i$ and $b_i$,
\be
\chi_z(x)=\exp(-\half\pi(x+nz)^2)\theta_u(y),
\ee
where $\theta_u(y)$ is holomorphic in the complex coordinates $y_i$.
Introducing isospin projection operators $I_{\pm}\equiv\half(1\pm\tau_3)$
one easily finds that
\be
D_z=\pmatrix{a_1&\hphantom{-}a_2^\dagger\cr a_2&-a_1^\dagger\cr}\!\otimes\!I_+
+\pmatrix{b^\dagger_1&\hphantom{-}b_2\cr b^\dagger_2&-b_1\cr}\!\otimes\!I_-.
\ee
With $D^\phd_zD^\dagger_z=(a_i^\dagger a_i+4\pi)\otimes I_+
+(b_i^\dagger b_i+4\pi)\otimes I_-$, one finds that the cokernel of $D_z$ is
trivial such that the Nahm transformation is well defined. The splitting of
$D_z$ in isospin-up and down components is a direct consequence of the fact
that $A$ is a reducible connection. The associated line bundle $L$ has a
section $s(x)=\chi_z(x)$ and the cocycle condition $\Psi_z(x+\lambda)=
g_\lambda(x) \Psi_z(x)$ reduces to
\be
\chi_z(x+\lambda)=\alpha(\lambda)\exp(-\half\pi i\lambda_\mu n_{\mu\nu}x_\nu)
\chi_z(x),
\ee
which implies that $\theta_u(y+q_c)$ equals $\theta_u(y)$ up to a $q$ and
$z$-dependent holomorphic factor, as required for holomorphic $\theta_u(y)$.
With $q_c$ we indicate the complex two-vector constructed form $q\in\zahlen^4$
as in eq.~(19) (cmp. $x_c=y$ and $(nz)_c=2iu$). From the general theory on
$\theta$-functions one finds that $\dim\,\ker\, a_i=1$, such that $\chi_z(x)$
is unique up to a ($z$-dependent) factor. For other choices of $n_{\mu\nu}$ it
can be easily shown that there are $\sqrt{\det(\half n)}$ such functions, as
required by the index theorem, see ref.~\cite{con} and references therein.
Explicitly
\bea
\chi_o(x)&=&\sum_{q\in\zahlen^4}\alpha(q)e^{-\half\pi \{ixnq+(x+q)^2\}},\\
\theta_o(y)&=&\sum_{q\in\zahlen^4}\alpha(q)e^{2\pi y\bar q_c-\half\pi
q^2},\nonumber
\eea
from which we can construct a smooth family of zero-modes
\be
\chi_z(x)=e^{-\pi ixz}\chi_o(x+\half nz).
\ee
Its norm is independent of $z$ and found to be~\cite{con} $\chi_o(0)
=(1.66925368)^2$, obtained by working out the sum over $\zahlen^4\times
\zahlen^4$ in $\int_{T^4}\chi^*_z(x)\chi_z(x)$, such that one of the sums
allows one to extend the integral over the unit cell to an integral over
$\real^4$, and evaluate the gaussian integral over $x$.

We therefore find two normalized zero-modes $\Psi_{ab}^{(i)}(x;z)=<a,b,x|
\Psi_z^{(i)}>$ for the Weyl operator $D_z$. The spinor and isospin indices
are denoted by $a$ and $b$. The  only non-zero components are
\bea
\Psi_{11}^{(1)}(x;z)&=&h(z)\chi_z(x)/\sqrt{\chi_o(0)},\nonumber\\
\Psi_{22}^{(2)}(x;z)&=&h^*(z)\chi^*_z(x)/\sqrt{\chi_o(0)},
\eea
where $h(z)$ is an arbitrary phase. This phase ambiguity is equivalent to a
gauge transformation in the subgroup generated by $\sigma_3$ (assigning a
different phase factor to $\Psi^{(2)}_z$ will introduce in addition a U(1)
gauge component, generated by $i\sigma_0$). We will choose $h(z)=1$.

One easily shows that eq.~(25) holds\footnote{
It should be noted that there are many representations of the $\theta$-function
by a lattice sum. Any holomorphic function $r(y)$, under some mild conditions
that guarantee convergence of the lattice sum, gives a proper solution by
defining $\tilde\theta(y)=\sum_{q\in\zahlen^4}\alpha(q)e^{2\pi y\bar q_c
-\half\pi q^2}r(y\!+\!q_c)$. For example, this leads to the following
alternative choice for a family of solutions, $\tilde\chi_z(q)=\sum_{q\in
\zahlen^4}\alpha(q)e^{-\half\pi\{ixnq+(x+q+nz)^2\}}$. The normalized solution
$\tilde\chi_z(x)/\tilde\chi_z(\!-\!nz)^\half$ is necessarily related to
$\chi_z(x)/\chi_o(0)^\half$ by a phase $h(z)$. As $\tilde\chi_z(-nz)$ vanishes
whenever $z_0\!=\!z_3\!=\!\half$ or $z_1\!=\!z_2\!=\!\half$, $h(z)$ represents
a singular gauge transformation, also signaled by the fact that $\tilde\chi_z$
gives rise to a trivial cocycle, $\tilde S_\lambda(z)=1$. Yet, with
$\hat{\tilde\omega}\!=\!\quart i\partial_\mu\log\tilde\chi_z(\!-\!nz)n_{\mu\nu}
dz_\nu$ one does retrieve the correct curvature, using that $\partial_{u_i}
\partial_{\bar u_i}\log\tilde\chi_z(\!-\!nz)=-2\pi\delta_{ij}$.}, and that
furthermore (note that $\alpha(\half nq)=\alpha(q)$)
\be
\chi_{z+q}(x)=\alpha(q)e^{-2\pi i qx}e^{-\half\pi i qnz}\chi_z(x).
\ee
{}From this one easily deduces that (see eq.~(8))
\be
S_\lambda(z)=\alpha(\lambda)\exp(-\half\pi i\lambda_\mu n_{\mu\nu}z_\nu\tau_3).
\ee
As $S_\lambda(z)=\hat g^{-1}_\lambda(z)$, we see that the cocycle for the
Nahm bundle is the inverse of the original cocycle (see eq.~(18)). The
topological charge is not affected by this inversion. This can also be
seen from the explicit computation of the connection $\hat\omega$ and
its curvature $\hat\Omega$. Using the fact that
\be
\partial_\mu\chi_o(x)=\half\pi in_{\mu\nu}x_\nu\chi_o(x),
\ee
it is almost trivial to show that
\be
\hat A_\mu(z)=-A_\mu(z),\quad \hat F_{\mu\nu}(z)=\tilde F_{\mu\nu}(z).
\ee

We could have anticipated the fact that the curvature of the Nahm bundle
is associated to the dual of (and hence minus) the curvature of the original
reducible bundle. The line bundle $L$ of its reducible U(1) component has
topological charge 1, determined from the square of the first Chern class,
$c_1(L)=\quart n_{\mu\nu}dx_\mu\wedge dx_\nu$. The above construction has
given us the Nahm transformation for this line bundle. Therefore we can
compute $c_1(\hat L)$ from eq.~(11), which is seen to give the desired result,
$c_1(\hat L)=\quart\tilde n_{\mu\nu}dz_\mu\wedge dz_\nu$.

The generalization to arbitrary constant curvature connections is now
straightforward and is left to the reader.

\section{Numerical results}

To warm up for our discussion of instantons on $T^3\times\real$ it is useful
to first consider the case of instantons for the O(3) non-linear sigma model
in two dimensions. The model is described by a field that lives on $S^2$,
parametrized through stereographic projection by a complex function u(x+iy).
Instantons~\cite{bel} on $\real^2$ are meromorphic functions
$u(z)=c\prod_i(z-a_i)/(z-b_i)$, where the topological charge equals the
number of poles. A standard result in complex function theory says
that a periodic meromorphic function with only one pole cannot exist,
ruling out the existence of charge 1 instantons on $T^2$. Higher charged
instantons are constructed from the Weierstrass $\sigma$-functions,
see ref.~\cite{ric}.

{}From the Hamiltonian point of view, the classical vacuum is degenerate along
the constant functions on the circle $S^1$. Its moduli space (the equivalent
of the space of flat connections on $T^3$) is equal to $S^2$, parametrized
by a complex number. Charge 1 instantons, corresponding to vacuum to vacuum
tunnelling events do, however, exist for $S^1\times\real$. Taking $S^1=
\real/2\pi\zahlen$ and introducing $z\equiv t+ix$ one finds~\cite{wip,sni}
\be
u(z)=-(c+de^z)/(a+be^z),\quad ad-bc\neq0.
\ee
It follows that $u(t\!\!\rightarrow\!\!\infty)\!=\!-d/b$ and $u(t\!\!
\rightarrow\!\!-\infty)\!=\!-c/a$, two {\em different} points in the
classical vacuum. As for $\real^2$ the charge 1 instanton has six parameters,
e.g. the complex ratios $b/a$, $c/a$ and $d/a$, two of which are associated
to translations, and three to O(3) rotations. The scale parameter is no longer
associated to a symmetry (as $S^1$ has a fixed size). A suitable
parametrization of charge one instantons~\cite{sni} is found in terms of the
polar angle $\fie\in[0,\pi]$
\be
u(z)=\ctg(\half\fie)-e^z/\sin(\half\fie),
\ee
chosen such that $u(t\!\!\rightarrow\!\!\infty)\!=\!\infty$ (the north pole)
and $u(t\!\!\rightarrow\!\!-\infty)\!=\!\ctg(\half\fie)$, whereas $u(0)$ is
real and the maximum of the energy occurs at $t=0$. This is seen to fix all
symmetries. Other solutions can be found by applying symmetry transformations
to this basic solution. Apparently the angle $\fie$ is related to the size of
the instanton.
On the other hand, $\fie$ is the geodesic distance between the points
$u(t\rightarrow\pm\infty)$ in the classical vacuum. A suitable definition
for the size $\rho$ of the instanton is given by~\cite{sni} $\rho(\fie)=
\sin(\half\fie)$. We explicitly see how the instanton is becoming singular
if one tries to impose periodic boundary conditions, $\fie\rightarrow0$. On
the other hand for antipodal $u(t\rightarrow\pm\infty)$, which are like
anti-periodic boundary conditions, the instanton has maximal size.
In this limit the energy profile is constant on $S^1$ and
the configuration at $t=0$ corresponds to a static solution of the equations
of motion with one unstable mode in the direction of tunnelling. This
configuration is therefore a sphaleron, which gives the minimal energy barrier
to be taken along the tunnelling path.

The situation for $T^3\times\real$ is seen to be quite similar. In the
absence of exact results we have used the lattice approximation to study
the instanton solutions numerically. Within the theory of finite volume
gauge theories~\cite{kol} these instanton solutions provide an essential
ingredient to go to larger volumes. Typically instantons become dynamically
relevant for the low-lying glueball spectrum in volumes around one half
to one cubic fermi. The finite volume sphaleron~\cite{mar} sets
the energy scale beyond which tunnelling effects are no longer exponentially
suppressed. The tunnelling path through the sphaleron~\cite{oim} gives
important information on the degrees of freedom that need to be accounted
for non-perturbatively. For a recent review of the dynamical aspects
see ref.~\cite{tre}.

The standard Wilson action for the lattice is a discrete approximation
to the continuum action. This discretization breaks the scale invariance,
such that the action of approximate instanton solutions will weakly depend
on the moduli parameters. In particular the action is decreasing as the scale
parameter of the instanton is lowered. As a consequence, looking for solutions
by lowering the action towards a local minimum (called cooling~\cite{tep}),
moves the configuration to (approximate) instantons of ever smaller size, at
some point ``falling through the lattice''. To reverse the shrinking of
instantons under cooling we have introduced over-improved cooling~\cite{oim}.
This is cooling with a modified lattice action, which reverses the size of
the scaling violations (rather than removing them to second order in the
lattice spacing). This is achieved by adding to the standard Wilson action
plaquettes of size $2\times2$, with a relative coefficient of $-1/40$. As a
consequence, the cooling now increases the size of the instanton, until it
cannot grow further due to the finite volume. Even on a lattice this provides
an exact solution. By scaling-up the number of lattice points a rather good
approximation to the continuum solution can be found, because lattice
artefacts are small for large instantons.

\begin{figure}[htb]
\vspace{9.5cm}
\includegraphics{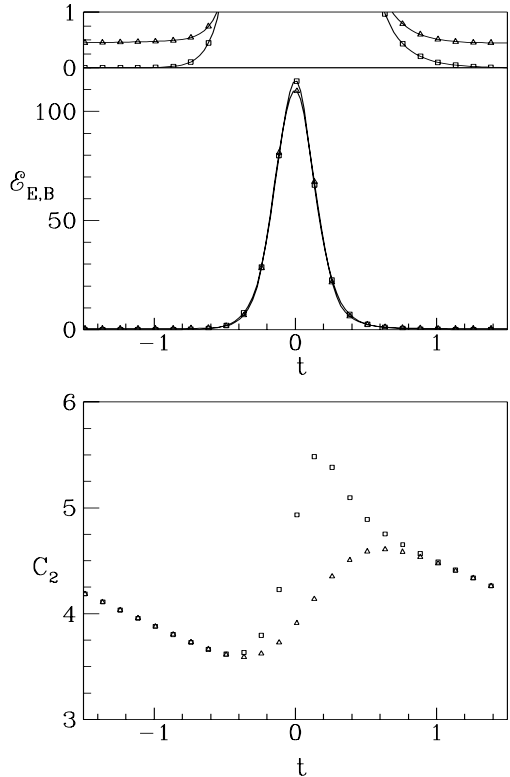}
\caption{Numerical results~[12] (after scaling appropriately with $N_s$) for
the case of an $8^3\times 24$ lattice with periodic boundary conditions,
obtained from over-improved cooling. In the top figure electric ($\cE_E(t)$
triangles) and magnetic ($\cE_B(t)$ squares) energies are plotted. The
tails are plotted at an enlarged scale in the inset. In the lower figure
we plot $C_2(t)$ through two distinct spatial points on the lattice.}
\end{figure}

We have clearly observed that the instanton does not like to have
periodic boundary conditions in the time direction. As we discussed
before, the obstruction against the existence of charge one instantons
on $T^4$ seems to imply that the configuration becomes peaked in the
limit that it tends to a solution. For the O(3) model this persists
even when time has an infinite extent. With the over-improved action
this effect is countered by the tendency of instantons to grow under
cooling. Indeed we found these effects to balance each other at some point.
Nevertheless, the resulting lattice solution clearly shows deviations from
self-duality, see fig.~1.

As we already described in the introduction, the classical vacua on $T^3$ are
given by flat connections. They can be parametrized by abelian constant vector
potentials $A_i^\flt=iC_i\tau_3/L$ for a three-torus of size $L$.
Not all of these are gauge inequivalent, and it is most convenient to
parametrize these flat connections by the Wilson loops that close because of
the periodic boundary conditions (also called Polyakov loops) for each of
the three generating circles of the torus
\bea
P_i(x)&\equiv&\half\Tr\Pexp\{i\int^L_0 A_i(x+s\hat e_i)ds\}\nonumber\\
&\equiv&\cos(C_i(x)/2).
\eea
One easily finds that $P^\flt_i(x)=\cos(C_i/2)$. For large extensions $T$
in the time direction, the finite action of an instanton requires the potential
energy to go to zero at both ends, such that $A_i$ will approach a flat
connection. As cooling brings the configuration to a solution of the equations
of motion, one easily finds that restricted to the set of flat connections, the
electric field ($\partial_t C_i$) has to vanish too. Self-duality would
require $C_i$ to become time-independent. From the lower part of figure 1 it is
clear that this is not the case for our lattice solution. Indeed, it can be
verified that the slope of $C_i(t)$ (only shown for $i=2$) is entirely
responsible for the electric energy in the tail regions where the magnetic
energy vanishes to a high accuracy. It should be noted that for increasing
$T$ ($N_t$ lattice units) this so-called electric tail will go down as
$1/T^2$ and from these numerical studies we can not rule out if instantons
do exist when $T\!\!\rightarrow\!\!\infty$.

\begin{figure}[htb]
\vspace{9.5cm}
\includegraphics{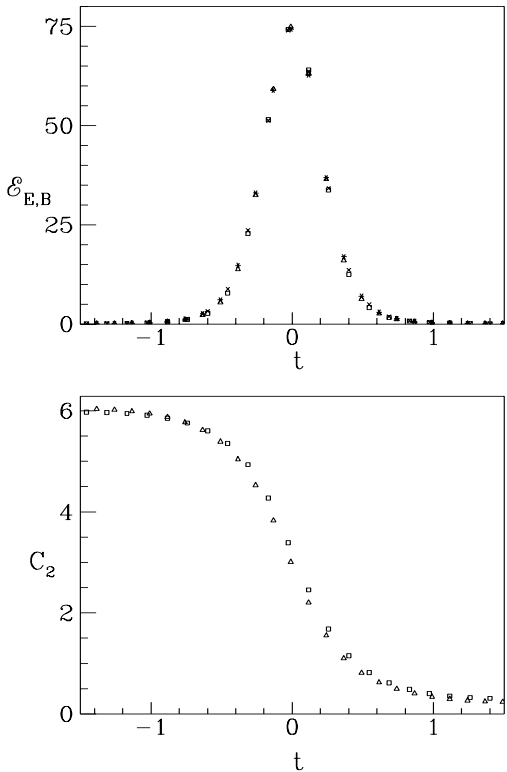}
\caption{Numerical results~[12] for lattices with $N_s=7$ and 8 (and
$N_t=3N_s$), properly scaled with $N_s$ and using twisted boundary conditions
($n_{0i}=1$). In the top figure electric ($\cE_E(t)$ squares at $N_s=7$ and
triangles at $N_s=8$) and magnetic ($\cE_B(t)$ crosses at $N_s=7$ and stars
at $N_s=8$) energies are plotted. In the lower part we plot $C_2(t)$ through
a given spatial point.}
\end{figure}

Nevertheless, the situation is dramatically different when twisting the
boundary conditions in the time-direction, as demonstrated in fig.~2. Lattice
artefacts are seen to be very small by comparing results on lattices with
$N_s=7$ and $N_s=8$. The electric tail has completely disappeared, as can also
be seen from the behaviour of $C_i(t)$, and the solution is perfectly
self-dual. Twisting the boundary conditions can be most simply achieved in
the $A_0=0$ gauge by applying an anti-periodic gauge transformation~\cite{tho},
\be
g(x)=\exp\{\pi i (x_1\!+\!x_2\!+\!x_3)\tau_3/L\}
\ee
at $t\!=\!T$, properly transposed to the lattice. It has the effect of
changing the sign of all Polyakov loops, $P_i(t\!=\!T)\! =\!-P_i(t\!=\!0)$.
One can also apply a twist such that, say, only $P_1$ is
(anti-)periodic. In all these cases instantons can be shown to exist. Assuming
that solutions persist under taking the limit $T\!\!\rightarrow\!\!\infty$,
this shows existence of instanton solutions on $T^3\times\real$.

It is tempting to expect that solutions will exist for any value of $P_i^{\rm
flat}$. It would give a natural way of counting the moduli parameters,
adding a scale parameter and the four translation parameters. It should
be said that this is not the way one counts the parameters when glueing in
a localized solution from $\real^4$ to the flat four-dimensional connection.
In this case there are three so-called attachment parameters~\cite{tau,mac}.
Assuming the conjecture to be true, however, we can prove the existence of
an instanton solution on $T^3\times\real$ with $P_i=0$ at {\em both} ends. We
have shown that this solution cannot be compactified, as it is also compatible
with periodic boundary conditions. It might, however, show the same behaviour
on $T^3\times\real$ in approaching $P_i(t\!\!\rightarrow\!\!\pm\infty)=0$ as
the O(3) instanton for $\fie\!\!\rightarrow\!\!0$. Over-improved cooling
with twisted boundary conditions will lead to a solution with a fixed value
of $P^\flt_i$, driven there by the well defined (but hard to control)
scaling violations.

Fixing $P_i^\flt$ at the the two ends is easily implemented on the
lattice. For the continuum, in the $A_0=0$ gauge, one would need to apply
a gauge transformation with unit winding number to the constant abelian
representation of the flat connection at $t=T$ to admit an instanton. On
the lattice a smooth configuration can only be specified by requiring gauge
invariant quantities to be smooth. A quantity like the vector potential
(on the lattice encoded in link variables) is in general not smooth, unless
one imposes in addition a ``smoothing'' gauge condition, like the Coulomb
gauge.
It implies that configurations can be in a singular gauge, with the singularity
``hidden'' between the meshes of the lattice. In this way the periodic boundary
conditions of the lattice are no obstacle for having smooth instanton
configurations.

Initially, fixed boundary conditions were introduced to search for instantons
with minimal energy barrier to locate the sphaleron\footnote{The sphaleron
was also constructed more directly using saddle-point cooling~\cite{mar}. It
was found to have the remarkable property that the gauge invariant part of the
magnetic field, $\Tr B_k^2(\vec x)$, is independent of the direction $k$ for
each point $\vec x$ separately. Cubic invariance only requires this to be the
case for the average. So far we have been unable to find an analytic expression
for this sphaleron.}. We can, however, now also study solutions with $|P_i|$
different at both ends. Our results indicate that, like for the case of
periodic boundary conditions, solutions exist that become self-dual when $T$
is chosen sufficiently large. If this is also true in the continuum remains
to be seen, as we have insufficient control over interchanging the limits
$N_t$ and $N_s\!\!\rightarrow\!\!\infty$.

\begin{figure}[htb]
\vspace{7cm}
\includegraphics{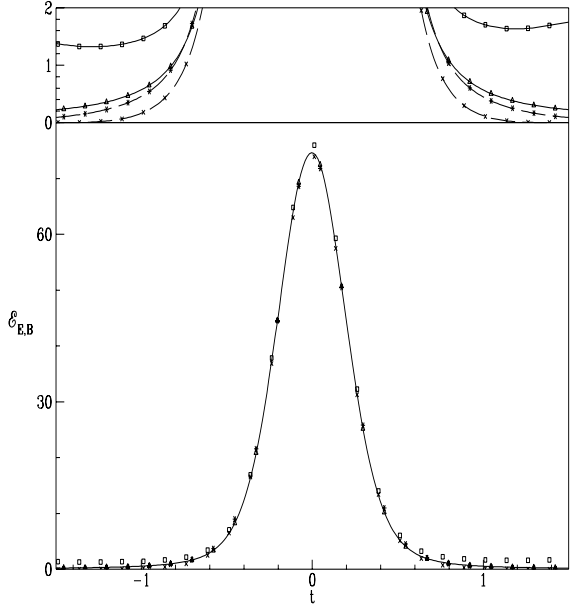}
\caption{
The electric and magnetic energies obtained after over-improved cooling
for a lattice of spatial size $N_s=8$ with boundary conditions fixed
in the time direction to $P_i=-1$ at one and $P_i=1$ at the other end~[13].
Results are for $N_t=24$ (squares for $\cE_E$ and crosses for $\cE_B$,
respectively the upper and lower curves in the inset) and $N_t=48$
(triangles for $\cE_E$ and stars for $\cE_B$, respectively the middle
upper and lower curves in the inset).}
\end{figure}

Since counting the number of moduli parameters (based on index theorems) in
general requires compact four-dimensional manifolds, it is not ruled out that
the moduli space of charge one instantons on $T^3\times\real$ is of higher
dimension than eight. We should point out, however, that despite of the
similarities with the non-linear sigma model, there is a marked difference.
Different points in the vacuum valley are not related by a symmetry. Indeed,
the shape of the static potential $V(A)$ depends strongly on the position
within the vacuum valley.  Our analysis has, however, established that the
instanton that corresponds to tunnelling through the sphaleron is associated
to $P_i(t\!\!\rightarrow\!\!\pm\infty)=\pm1$. These are rather special points
of the vacuum valley, with directions in field space where $V(A)$ shows
quartic, rather than quadratic behaviour. This also means that the approach
of the tunnelling path to these points in the vacuum valley is not guaranteed
to be exponential in time, as seems to be confirmed by our numerical results
illustrated in fig.~3. Note that by forcing the configuration to be in the
vacuum valley at either end (with $P_i^\flt=\pm1$) we also destroy the
self-duality, in spite of the fact that the configuration in fig.~3 is
compatible with twisted boundary conditions. At finite $T$, instantons do no
longer correspond to vacuum-to-vacuum tunnelling solutions (it takes an
infinite time to exactly reach the vacuum).

It might very well be that in the continuum only instantons with $P_i(t\!\!
\rightarrow\!\!\pm\infty)=\pm1$ will be exact solutions. At the other extreme
there seems to be the possibility of having three additional moduli parameters.
{}From the physical point of view the most urgent problem would be to get
some analytic hold on the instanton solution that tunnels through the
sphaleron, so as to be able to take these effects into account beyond a
semiclassical approximation, mandatory when the wave functionals of states
that appear in the low-lying spectroscopy will no longer be exponentially
suppressed near the sphaleron. For $S^3\times\real$ such a study~\cite{sth}
has been recently shown to be feasible~\cite{heu}. Also from the mathematical
point of view, the properties of the moduli space for $T^3\times\real$ presents
an interesting challenge. In the next section we make some general observations
about the Nahm transformation in this context.

\section{The Nahm transformation on $T^3\times\real$}

When time extends to infinity, the partial integration involved, going
from eq.~(13) to (16) can pick up a boundary term,
\bea
\hat F_{\mu\nu}^{ij}(z)\!\!\!\!&=&\!\!\!\!8\pi^2<\Psi^{(i)}_z|\sigma^\dagger_{
[\mu}\sigma^\phd_{\nu]}G_z|\Psi_z^{(j)}>\\&-&\!\!\!\!4\pi
i\oint\!\frac{\partial
\Psi^{(i)}_z(x)^\dagger\!\!}{\partial z_{\,[\mu}}\,\sigma^\dagger_\lambda
\sigma^\phd_{\nu]}\!\left(\!G_z\Psi_z^{(j)}\!\right)\!(x)\,d^3_\lambda x.
\nonumber
\eea
This boundary term in general destroys the anti-selfduality of $\hat\Omega$.
It is useful to review the situation for $\real^4$, so as to get some insight
in how the Nahm transformation needs to be modified in the presence of these
boundary terms~\cite{nam,cor}. If $\Psi(x)$ is a normalized zero-mode of the
Weyl operator $D$, also $e^{-2\pi i zx}\Psi(x)$ is normalized and easily
seen to be a zero-mode of $D_z$. Hence
\bea
\Psi_z(x)&=&e^{-2\pi i zx}\Psi(x),\\
G_z(x,y)&=&e^{2\pi iz(y-x)}G(x,y).\nonumber
\eea
Compactifying to $S^4$ and using conformal invariance allows for a regular
description of the gauge field at infinity, after a suitable gauge
transformation $g(x)$ (with unit winding number). From this a solution of
the Weyl equation is seen to have the following asymptotic
behaviour~\cite{nam,cor}
\bea
\Psi(x)\!\!\!\!&=&\!\!\!\!\sigma^\dagger_\mu x^\phd_\mu g(x)\frac{\alpha}{\pi}
|x|^{-4}\!+\!\cO(|x|^{-4}),\\(G\Psi)(x)\!\!\!\!&=&\!\!\!\!-\quart
\sigma^\dagger_\mu x^\phd_\mu g(x)\frac{\alpha}{\pi}|x|^{-2}\!+\!\cO(|x|^{-3}),
\nonumber
\eea
where $\alpha$, associated to each zero-mode, is a constant factor with spinor
and isospin (i.e. group) indices. We easily deduce from this that the
Nahmtransformed connection is $z$-independent and that its curvature is no
longer anti-selfdual
\bea
\hat A^{ij}_\mu\!\!\!\!&=&\!\!\!\!\!-2\pi i<\!\Psi^{(i)}|x_\mu|\Psi^{(j)}\!>,\\
\hat
F_{\mu\nu}^{ij}\!\!\!\!&=&\!\!\!8\pi^2\!<\!\Psi^{(i)}|\sigma^\dagger_{[\mu}
\sigma^\phd_{\nu]}G|\Psi^{(j)}\!>\!+\pi^2\alpha^\dagger\sigma^\phd_{[\mu}
\sigma^\dagger_{\nu]}\alpha.\nonumber
\eea
The boundary term in this case has added a self-dual contribution to the
anti-selfdual part that was already present. Nevertheless, one can proceed
to attempt to perform the Nahm transformation a second time. The Weyl operator
now becomes algebraic, $\hat D^-_x\!=\!\sigma_\mu(\hat A_\mu\!+\!2\pi i
x_\mu)$,
but $\hat D^-_x\hat D^+_x$ no longer commutes with the quaternions, since
$\hat F_{\mu,\nu}=[\hat A_\mu,\hat A_\nu]$ has a self-dual component. In this
way, applying the Nahm transformation again will not give an anti-selfdual
connection. However, one can modify the Weyl operator as follows~\cite{nam,cor}
\be
\hat{\tilde D}^-_x\!\equiv\!\left(\!2\pi\sigma_2\alpha,\!\hat D_x^-\!\right)\!=
\!\left(\!2\pi\sigma_2\alpha,\sigma_\mu(\!\hat A_\mu\!+\!2\pi i x_\mu\!)\!
\right)\!.
\ee
This leads to the desired commutation with the quaternions
($\tr$ is the trace on the spinor indices)
\bea
\hat{\tilde D}^-_x\hat{\tilde D}^+_x\!\!&=&\!\!4\pi^2\alpha\!\otimes\!
\alpha^\dagger\!+\!(\hat A_\mu\!+\!2\pi i)^2\!+\!\sigma^\phd_{[\mu}
\sigma^\dagger_{\nu]}\hat F_{\mu\nu}\nonumber\\{}\!\!&=&\!\!(\hat A_\mu\!+
\!2\pi_i)^2\!-\!2\pi^2\tr(\alpha\!\otimes\!\alpha^\dagger),
\eea
using $\sigma^\phd_{[\mu}\sigma^\dagger_{\nu]}\hat F_{\mu\nu}\!=\!\sigma^\phd_{
[\mu}\sigma^\dagger_{\nu]}\alpha^\dagger\sigma^\phd_{[\mu}\sigma^\dagger_{\nu]}
\alpha\!=\!-4\tau_i\alpha^\dagger\tau_i\alpha$ and the completeness relation
for the Pauli matrices, $\tau_i^{ab}\tau_i^{cd}\!=\!\delta_{ad}\delta_{bc}\!-
\!\half\delta_{ab}\delta_{cd}$. The Nahm transformation can be shown to be
complete and coincides with the algebraic ADHM~\cite{adhm} construction.

It is also worthwhile to discuss the construction for monopoles~\cite{nam},
which are time independent self-dual solutions on $\real^3$. Here $A_0$ plays
the role of the Higgs field, required to approach a fixed length at infinity,
which can always be scaled to one. For SU(2) we therefore can choose
$A_0\!\!\rightarrow\!\!-\half i g(x)\tau_3 g^{-1}(x)$, for $r\!=\!|\vec x|
\!\!\rightarrow\!\!\infty$, with $g(x)$ a non-trivial gauge transformation.
As for the instantons, one easily sees that $\hat A_\mu(z)$ is independent
of $z_i$. The remaining parameter $z\!=\!z_0$ would be expected to have
infinite range. But for $\Psi\in\ker D_z$, the asymptotic behaviour of the
Weyl equation is seen to require $\det(A_0+2\pi_i z)<0$, in order for it to
to admit normalizable zero-modes~\cite{nam,cal}, thus $z\!\in\!(\!-\pi,\pi)$.
Amazingly, the BPS~\cite{bps} monopole solution can be retrieved by putting
$\hat A_\mu(z)=0$ as a solution to the one dimensional duality equations
on this interval and by performing the Nahm transformation~\cite{nam}.

For $T^3\times\real$,  eq.~(37) will only give a non-trivial boundary term
from the $\lambda\!=\!0$ contribution. For convenience we will choose
$T^3=\real^3/\zahlen^3$, i.e. $L=1$. Like for $\real^4$, the $z_0$
dependence associated with the non-compact time parameter, becomes trivial,
$\Psi_z(x)=\exp(-2\pi i z_0 t)\Psi_{\vec z}(x)$. The Weyl equation, with
$D_{\vec z}$ its spatial part, reduces to
\be
D_z\Psi_z(x)=e^{-2\pi i z_0 t}(\partial_0+D_{\vec z})\Psi_{\vec z}(x),
\ee
in the $A_0\!=\!0$ gauge, and we can study the zero-modes in terms of the
spectral decomposition of the spatial part of the Weyl equation, which for
a flat connection reduces, up to a gauge, to
\be
D_{\vec z}(A^\flt)e^{2\pi i\vec p\cdot\vec x}\Psi^s_{\vec z}(\vec p)\!
=\!e^{2\pi i\vec p\cdot\vec x}E_{\vec z}^s(\vec p)\Psi^s_{\vec z}(\vec p),
\ee
where $s=\pm\half$ is the isospin of the solution and $\vec p\in\zahlen^3$
gives the momentum. The momentum and isospin eigenstates satisfy the spinor
equation
\be
\tau_i(2\pi\vec p\!+\!\half s\vec C\!+\!2\pi\vec z)\Psi^s_{\vec z}(\vec p)\!
=\!E_{\vec z}^s(\vec p)\Psi^s_{\vec z}(\vec p),
\ee
with positive and negative energy eigenvalues given by
\be
E_{\vec z}^{s}(\vec p)=\pm\|2\pi\vec p+s\vec C+2\pi\vec z\|.
\ee
When $A=A^\flt$, the zero-modes can be decomposed according to
\be
\Psi_{\vec z}(x)=\sum_{\vec p,s,v} a^{s,v}_{\vec p}(t)
e^{2\pi i\vec x\cdot\vec p}\Psi^{s,v}_{\vec z}(\vec p),
\ee
where $v$ labels the positive or negative energy states\footnote{In the
background of an instanton configuration which connects the flat connections
at either end, each eigenstate of $D_{\vec z}(\vec A(t,\vec x))$ evolves
smoothly as a function of $t$ from one state in the spectrum determined by
eq.~(46) at $t\!=\!-\infty$ to another at $t\!=\!\infty$, of course with the
appropriate values of $\vec C$ inserted at either end. From the relation
between the index theorem and spectral flow, we still expect that
generically there is exactly one such state $\Psi_t$ (with eigenvalue $E_t$)
that moves from negative energy at $t\!=\!-\infty$ to positive energy at
$t\!=\!\infty$. Eq.~(43) would imply that in an adiabatic approximation
$\Psi_{\vec z}\sim\exp(-E_tt)\Psi_t$ and the spectral flow provides a natural
explanation for the existence of a normalizable zero-mode. We will not attempt
to make this more precise here.}.

At fixed $\vec C$, for all but a finite number of values of $\vec z$ within
one unit cell of $\hat T^3=\real^3/\zahlen^3$, the spectrum of $D_{\vec z}
(A^\flt)$ has a gap, in which case the zero-modes of the Weyl equation are
expected to decay exponentially in time. There would in this case be no
boundary term contributing to eq.~(37). As long as $\vec C\!\neq\!\vec 0\mod
2\pi$, any instanton solution will approach $A^\flt$ exponentially in time
and the conclusion that $\Psi_{\vec z}$ decays exponentially in time will
hold. If, however, $A^\flt$ is associated to one of the quartic points in
the vacuum valley, $\vec C\!=\!\vec 0\mod2\pi$, this might require
more care. Assuming for the moment that this will cause no problems for the
behaviour of the zero-modes of the Weyl equation, also in this case
boundary terms will be absent for $\vec z\!\neq\!\vec 0\mod\pi$. As a
consequence eq.~(16) will remain valid almost everywhere, with a correction
that has support at a finite number of points only. From the theory of
distributions, this correction term should be expressible in terms of delta
functions. This provides the interesting suggestion to study the BPS
equations, $B_k=D_k A_0$, in the presence of point-like source.
Performing the Nahm transformation away from these finitely many point-sources
might provide us with a method of constructing instanton solutions
on $T^3\times\real$. Future work will tell if we require more detailed
information on these singularities in order to successfully construct
non-trivial solutions.

\section{Conclusion}

We have reviewed the role of the Nahm transformation in constructing solutions
of the self-duality equations for gauge theories, with an emphasis on the
applications to $T^4$ and in particular $T^3\times\real$. On $T^4$ we
explicitly illustrated this transformation for the class of (reducible)
constant curvature solutions, which form special (singular) points in the
moduli space. The quest for explicit instanton solutions on a torus is
motivated by the dynamical study of glueball spectroscopy in small to
intermediate volumes. Numerical lattice studies of these instantons have
therefore been developed recently and have been reviewed here too. These
numerical studies have led to interesting conjectures on the moduli space
of charge one instantons on $T^3\times\real$. It provides the motivation for
a renewed attack on the study of the Nahm transformation for this setting. We
showed how singularities arise due to certain boundary terms (like for
$\real^4$), related to the asymptotic behaviour of the chiral zero-modes
of the Dirac-Weyl equation. The study of the BPS equations on $T^3$ with
finitely many sources is hoped to give us an analytic technique to
construct the charge one instanton on $T^3\times\real$, using the
Nahm transformation.

\section*{Acknowledgements}

I am grateful to Gerhard Weigt and Dieter L\"ust for their generous invitation
to the Buckow meeting. I thank Andreas Wipf for discussions concerning
instantons for the O(3) model. I would like to also thank my many collaborators
for their interest in these problems. In particular I am grateful to
Peter Braam and Margarita Garc\'{\i}a P\'erez for many stimulating and helpful
discussions over the years. The numerical aspects of this work
were supported in part by a grant from ``Stichting Nationale Computer
Faciliteiten (NCF)'' for use of the CRAY C98 at SARA.

\end{document}